\documentclass[12pt]{iopart}  %aip,jmp     aps,prb
\usepackage{graphicx}%
\usepackage{bm}% 
\usepackage{bbold}
\usepackage[normalem]{ulem}
\usepackage[colorlinks=true,linkcolor=blue,citecolor=blue,urlcolor=blue]{hyperref}
\usepackage{xcolor}
\usepackage[margin=1.5cm]{geometry}
\usepackage{cite}
\usepackage{pstricks,pst-node}
\usepackage{amsfonts}

\begin{document}

\newcommand{\op}[1]{{\bm{#1}}}
\newcommand{\bra}{\langle}
\newcommand{\ket}{\rangle}

\newcommand{\new}[1]{\textcolor[rgb]{0,0,1}{#1}}

\newcommand{\old}[1]{\textcolor[rgb]{0.5,0.5,0.5}{\sout{#1}}}
\newcommand{\nota}[1]{\textcolor[rgb]{1,0,0}{{#1}}}
\newcommand{\Oplus}{\ensuremath{\vcenter{\hbox{\scalebox{1.5}{$\oplus$}}}}}

%\renewcommand{\multirowsetup}{\centering}
%\newlength{\LL}\settowidth{\LL}{Atomic Levels} 
%************************

\title{From free-evolution to tomographic representation}
\author{S. Cordero$^{1}$ R. L\'opez-Pe\~na$^{1}$, E. Nahmad-Achar$^{1}$, O. Casta\~nos$^{1}$, J. A. L\'opez-Sald\'ivar$^{2,3}$, V. I. Man'ko$^{4}$} 

%\email{sergio.cordero@nucleares.unam.mx}
%
%%\email{lopez@nucleares.unam.mx}
%%
%\author{E. Nahmad-Achar}
%%\email{nahmad@nucleares.unam.mx}
%%
%\author{O. Casta\~nos}
%%\email{ocasta@nucleares.unam.mx}
%%
%\author{R. L\'opez-Pe\~na}
%%\email{lopez@nucleares.unam.mx}
%%
%\author{V. I. Man'ko}
%%\email{manko@sci.lebedev.ru}
%%
%\author{J. A. L\'opez-Sald\'ivar}
%%\email{julio.lopez.8303@gmail.com}
%%

\address{
$^{1}$Instituto de Ciencias Nucleares, Universidad Nacional Aut\'onoma de M\'exico, Apartado Postal 70-543, 04510  Mexico City, Mexico\\

$^{2}$ Russian Quantum Center, Skolkovo, 143025 Moscow, Russia \\

$^{3}$ Laboratory of Quantum Information Technologies, National University of Science and Technology “MISIS”, 119049 Moscow, Russia \\

$^{4}$ Lebedev Physical Institute, Russian Academy of Sciences, Leninskii Prospect 53, 119991, Moscow, Russia.}

\date{\today}

\begin{abstract}
We use the free evolution propagator to determine the quantum probability representation (i.e., the general expression of the tomogram) of any one-dimensional system described by a density state. The evolution operator for the considered quantum system is additionally used to establish the corresponding time dependent tomogram. Applications are given for a Gaussian wave packet, the quantum shutter related with the phenomenon of diffraction in time, the double quantum shutter, and a finite potential. A generalisation to describe $N$ particle systems is also presented and, in particular, we find the tomogram associated to the 2 particle case occupying in general non-orthogonal states. In the latter case, for a bipartite quantum system, the entanglement properties are established by considering quantum information concepts such as the linear entropy. 
\end{abstract}

%
%\pacs{}%
%

\maketitle

\section{Introduction}
Generally, pure quantum states are associated with vectors in a Hilbert space while mixed states with state operators, which can be studied in the Schr\"odinger, Heisenberg and Dirac representations. Another representation closer to classical statistical mechanics is the phase space formulation of quantum mechanics which leads to the establishment of quasi-probability distribution functions like, for example, the Wigner~\cite{wigner32}, Husimi~\cite{husimi40}, and Glauber-Sudarshan functions~\cite{glauber63, sudarshan63}. This formalism allows one to calculate all the usual observables that the wave function approach estimates.  Later on, one these quasi-probability distribution functions was unified into a one-parameter family called s-ordered quasi-distribution functions~\cite{lee95}.  These distribution functions have been very useful in many areas of quantum physics with the purpose to visualising its behaviour in phase space and exploring the corresponding quantum effects.

They receive the name of ``quasi-distribution functions'' because they are not real probability distribution functions, i.e., The Wigner and P functions can have negative values, and the Q function does not describe measurable distributions of physical observables. However, in the 1990's the tomographic approach emerged where one can connect quantum states with real probability distributions. This formulation appeared first in the area of quantum optics, with the optical homodyne tomography, based on the formalism introduced by Glauber, Vogel and Risken where the Cahill-Glauber quasi probability distributions can be reconstructed by measuring probability distributions of quadrature variable.~\cite{glauber63, cahill69, vogel89}. It is an integral relation between the Wigner function and the marginal distribution for a measurable homodyne variable~\cite{vogel89}.  Later, the symplectic tomographic approach was established to get the Wigner function by measuring a real probability distribution function in a scaled and rotated reference frame~\cite{ibort09}, that is, the probabilities are directly measured in experiments. 

The entanglement in quantum physics provides the possibility to apply the classical probability theory~\cite{chernega23b}. The functions that define the states of a quantum system, named {\it symplectic tomograms}, constitute real probability distribution functions and were introduced in~\cite{mancini95,mancini96}. These symplectic tomograms have been also called {\it probability representation of quantum mechanics}, and are discussed in~\cite{mancini97, omanko97, przhiyalkovskiy22, asorey15}.

The tomogram may be read from the free evolution of an initial wave packet, via Green's function or otherwise, identifying appropriately the corresponding time and mass of the particle with the reference frame parameters. Dually, from the tomogram obtained from a quorum of reference frames one may reconstruct the free evolution of the initial wave packet.

The manuscript is organized as follows: In section~\ref{sec2} a review of the properties of the tomographic representation of quantum systems is presented and focused to its connection to the free propagator. The time-dependent representation is also obtained through the evolution operator. Using this connection, the symplectic tomogram of several quantum systems are obtained in section~\ref{sec3}, viz., a Gaussian wavepacket, the single and double shutter problems, and a finite multi-barrier potential. In section~\ref{sec4} a generalisation  to the multimodal harmonic oscillator in the tomographic representation is discussed, taking into account that a different reference system needs to be taken for each particle. Section~\ref{sec5} gives the application to the study of entanglement in a bipartite non-orthogonal system. A summary and some concluding remarks are given at the end.

\section{Tomographic representation}
\label{sec2}

The symplectic tomogram $\mathcal{W}(X \vert \mu, \nu)$ of the system can be obtained via the Wigner function
\begin{equation}
W(q,p)=\frac{1}{2\pi}\int_{-\infty}^{\infty} \left \langle x- \frac{\xi}{2} \right \vert \hat \rho \left \vert x+\frac{\xi}{2} \right \rangle e^{i p \xi} \, d\xi \,,
\end{equation}
where we have set $\hbar=1$, by a two-dimensional Fourier transformation: 
\begin{equation}
\mathcal{W}(X \vert \mu, \nu) = \int e^{-i k(X-\mu q - \nu p)}W(q,p) \frac{dk \, dq\, dp}{(2 \pi)^2} \,.
\end{equation}
The physical meaning of the parameters $\mu$ and $\nu$ is that they describe an ensemble of rotated and scaled reference frames in which the position $X$ is measured (cf.~\cite{manko99b}).

The density operator can be obtained from the tomogram as follows
\begin{equation}
\op{\rho} = \frac{1}{2 \pi} \int \mathcal{W}(X \vert \mu, \nu) e^{i(X \op{I}- \mu \op{x}\,-\nu \op{p})} dX \, d\mu \, d\nu,
\end{equation}
which can be used to obtain the Husimi probability distribution $Q_\rho( \alpha)=\langle \alpha \vert \op{\rho} \vert \alpha \rangle / \pi$, as
\begin{equation}
Q_\rho( \alpha) = \frac{1}{2 \pi^2}\int \mathcal{W}(X \vert \mu, \nu) e^{-\frac{1}{2}\vert z \vert^2 -i(\alpha z^* +\alpha^* z+X)} dX\, d\mu \, d\nu, \quad z=\frac{1}{\sqrt{2}}(\mu+i\nu).
\end{equation}

To determine the tomographic representation of common quantum mechanical wave functions one considers the family of observables~\cite{ibort09},
\begin{equation}
\op{X}(\mu, \nu) = \mu \, \op{Q} + \nu\, \op{P} \, , \quad  \op{Y}(\mu, \nu) = -s^{2} \nu  \, \op{Q} + s^{-2} \mu\, \op{P}\, ,
\end{equation}
which satisfy the commutation relations $[ \op{X},\op{Y}]=[ \op{Q},\op{P}]=  i \hbar$, with the condition $ (\mu/s)^2 + (\nu s)^2 =1$, where $s, \mu, \nu$ are real parameters. In this way one has a canonical transformation preserving the symplectic form in the phase space. Of course, the spectrum of the observables is continuous and generates a real line. In general the parameters $\mu, \nu$ describe an ensemble of rotated 
$(\phi)$ and scaled ($s$) reference frames in which the position $X$ or momentum $Y$ can be measured,
\begin{equation}
\mu = s \cos\phi , \quad \nu = s^{-1} \sin\phi \, .
\end{equation}
One can determine the eigenstate of the position operator $\op{X}$, denoted by $| X, \mu,  \nu \rangle$ for arbitrary values of $\mu$ and $\nu$, as
\begin{eqnarray}
\langle x | X, \mu, \nu \rangle &=& \frac{1}{\sqrt{2 \, \pi \hbar |\nu|} } e^{-i \frac{\mu x^2}{2 \hbar \nu} + i \frac{X x}{\hbar \nu}} \, ,  \nonumber \\
\langle p | X, \mu, \nu \rangle &=& \frac{1}{\sqrt{2 \, \pi \hbar |\mu|} } e^{i \frac{\nu p^2}{2 \hbar \mu} - i \frac{X p}{\hbar \mu}} \, ,
\end{eqnarray}
in the position and momentum representation, respectively. 

The amplitude of the probability density of an arbitrary pure state $| \psi\rangle$, in the rotated and scaled reference frame, is given by
\begin{equation}
\langle X, \mu, \nu | \psi\rangle = \int_{-\infty}^{\infty}\, \langle X, \mu, \nu | x^\prime \rangle \langle x^\prime| \psi \rangle {\rm d}x' \, .
\end{equation}
Therefore, the conditional probability may be written as follows
\begin{equation}\label{eq.Wxmunu}
\mathcal{W}_\psi( X | \mu ,\nu) = \frac{1}{2 \, \pi \hbar |\nu|} \Bigg| \int^\infty_{-\infty}\, \psi(x) \, e^{i \frac{\mu x^2}{2 \hbar \nu} - i \frac{X x}{\hbar \nu}} {\rm d} x \Bigg|^2  \, ,
\end{equation}
where $\nu \neq0$.

A similar procedure can be done in the momentum representation, and in this case the conditional probability takes the form
\begin{equation}\label{eq.Wpnumu} 
\mathcal{W}_\psi( X | \mu, \nu) = \frac{1}{2 \, \pi \hbar |\mu|} \Bigg| \int^\infty_{-\infty}\, \psi(p) \, e^{-i \frac{\nu p^2}{2 \hbar \mu} + i \frac{X p}{\hbar \mu}}  {\rm d}p \Bigg|^2 \, ,
\end{equation}
where $\mu \neq0$. Of course, $\psi(p)=\bra p|\psi\ket$ is related by the Fourier transform of $\psi(x)=\bra x|\psi\ket$ and the tomographic probability representation of the pure state is normalised.  Thus one is able to reconstruct the marginal probability densities $|\psi(x)|^2$ and $|\psi(p)|^2$ by means of a quorum of measurements of $X(\mu, \nu)$.  The same procedure can be applied for the density matrix $\rho(x,x^\prime) = \psi(x) \psi^*(x^\prime)$, and even for mixed states, as will be seen below.

\subsection{Tomographic representation as the free evolution}\label{s.freeU}

Manipulating the exponent inside the integral of equation~(\ref{eq.Wxmunu}), one may rewrite the tomogram as 
\begin{equation}\label{eq.Wpsimunu}
\mathcal{W}_\psi( X | \mu ,\nu) = \Bigg| \frac{1}{\sqrt{\mu}} \varphi(X;\mu, \nu) \Bigg|^2 \, ,
\end{equation}
with the probability amplitude $ \varphi(X;\mu, \nu)$ defined as
\begin{equation}\label{eq.varphi}
\varphi(X;\mu, \nu):=\sqrt{\frac{\mu}{2 \pi i\, \hbar \nu} } \, \, \int^\infty_{-\infty} \, \exp\left[i \frac{\mu}{2\hbar\nu}\left(\frac{X}{\mu}-y\right)^2\right] \psi(y) {\rm d}y\,. 
\end{equation}
Notice that the global phase $e^{i \frac{X^2}{2 \hbar \mu \nu}}$ was neglected because it does not change the tomogram. This expression reminds us of the propagator for a free particle of mass $m$ from position $y$ at time $\tau=0$ to position $x$ at time $\tau>0$, 
\begin{equation}\label{eq.G0xt}
G_0(x,\tau;y,0)=\sqrt{\frac{m}{2\pi i\,\hbar \tau}} e^{i m (x-y)^2/2\hbar \tau}\, ,
\end{equation}
so for values  $\mu>0$ and $\nu>0$, by simple comparison one identifies 
\begin{equation}
\frac{\nu}{\mu} = \frac{\tau}{m}\,,\qquad \frac{X}{\mu}=x\, .
\label{munu}
\end{equation}
By considering the free evolution of a wavepacket $\psi(y)$
\begin{equation}\label{eq.psixt}
\psi(x,\tau) = \int_{-\infty}^{\infty}\, G_0(x,\tau;y,0) \psi(y){\rm d} y =  \bra x|e^{-i\op{H}_0\tau/\hbar}|\psi\ket\, =: \bra x| \op{G}_0(\tau) |\psi\ket
\end{equation}
where $\op{H}_0$ is the free Hamiltonian of the system, one is able to write
\[
\varphi(X;\mu,\nu) \varphi^\ast(X;\mu,\nu) = \langle X,\mu,\nu \vert \op{G}_0 \vert\psi\rangle 
\langle \psi\vert \op{G}_0^\dagger \vert X,\mu,\nu \rangle
\]
and the tomogram (\ref{eq.Wpsimunu}) takes the form
\begin{equation}
\mathcal{W}_\rho( X | \mu ,\nu) = \frac{1}{\vert\mu\vert} \langle X,\mu,\nu \vert \op{G}_0 \op{\rho} \op{G}_0^\dagger \vert X,\mu,\nu \rangle
\end{equation}
with $\op{\rho} = \vert\psi\rangle\langle \psi\vert$. That is, given the density operator (even for mixed states), the tomogram may be obtained by taking the free-evolution operator on the left and right as shown above, with the identification given by~(\ref{munu}).

The non-negativity, integrability, and homogeneity of the tomogram~\cite{ibort09} are easily verified through eq.~(\ref{eq.G0xt}) evaluated at $-\tau$
\begin{equation}\label{eq.wpsixt}
\widetilde{\psi}(x,\tau) := \int_{-\infty}^{\infty} \, G_0(x,-\tau;y,0) \psi(y){\rm d}y 
= \left[\int_{-\infty}^{\infty}\, G_0(x,\tau;y,0) \psi^*(y){\rm d} y\right]^*\,,
\end{equation}
which allows us to establish the tomogram $\mathcal{W}_\psi( X | \mu ,\nu)$ for any value of $\mu, \nu$
\begin{eqnarray}
\mathcal{W}_\psi( X | \mu ,\nu) &=& \frac{1}{|\mu|} \left|\psi\left(\frac{X}{\mu}, \frac{m}{\mu}\nu\right) \right|^2 \,,\nonumber \\[2mm]
\mathcal{W}_\psi( X | -\mu ,\nu) &=&\frac{1}{|\mu|} \left| \widetilde{\psi}\left(-\frac{X}{\mu}, \frac{m}{\mu}\nu\right) \right|^2 \,,\nonumber \\[2mm]
\mathcal{W}_\psi( X | \mu ,-\nu) &=& \frac{1}{|\mu|} \left| \widetilde{\psi}\left(\frac{X}{\mu}, \frac{m}{\mu}\nu\right) \right|^2 \,,\nonumber \\[2mm]
\mathcal{W}_\psi( X | -\mu ,-\nu) &=& \frac{1}{|\mu|} \left|  \psi\left(-\frac{X}{\mu},  \frac{m}{\mu}\nu\right) \right|^2 \,,\label{eq.Wpsi-full}
\end{eqnarray}
where we are considering here $\mu,\,\nu>0$.
Notice that the dependency on $m\nu/\mu$ of the wave function yields a tomogram independent of the mass $m$.
Also, one is able to prove that the Green function~(\ref{eq.G0xt}) is invariant when the mass and the time are rescaled, i.e., 
\[G_0(x,\lambda \tau;y,0)\bigg|_{m\to \lambda m} = G_0(x,\tau;y,0)\,,\]
property that in the tomogram reads as $\mathcal{W}_\phi(\lambda X| \lambda\mu,\lambda\nu) = \mathcal{W}_\phi(X| \mu,\nu)/|\lambda| $.

The above shows that the free time-dependent solution $\psi(x,\tau)$, together with $\widetilde{\psi}(x,\tau)$, provides in a simple way the tomogram of a wavepacket. Notice that $\widetilde{\psi}(x,\tau) = \psi^*(x,\tau)$ for real initial states $\psi(y)=\psi^*(y)$. 

\subsection*{Limits}

To complete the analysis we consider the limits $\nu\to 0$ when $\mu\neq 0$, and $\mu\to 0$ when $\nu\neq 0$; both quantities being naught making no sense. Since $\nu/\mu = \tau/m$ one obtains, for $\mu\neq 0$
\[
\lim_{\tau\to 0^+} G_0(x,\tau;y,0)=\delta(x-y)\,;
\]
this establishes the equality $\lim_{\tau\to0^+} \psi(x,\tau) = \lim_{\tau\to0^+} \widetilde{\psi}(x,\tau) = \psi(x,0)$ for equations~(\ref{eq.psixt}) and (\ref{eq.wpsixt}), and hence
\begin{eqnarray*}
\mathcal{W}_\psi( X | \mu ,0) &=&  \frac{1}{|\mu|} \left| \psi\left(x, 0\right) \right|^2  \bigg|_{x=X/\mu}\,,\\[2mm]
\mathcal{W}_\psi( X | -\mu ,0) &=& \mathcal{W}_\psi( -X | \mu ,0)\,.\\
\end{eqnarray*}
In this limit the tomogram is proportional to the probability distribution in the position representation.
On the other hand, for the limit $\nu\to0^+$, one uses~(\ref{eq.Wpnumu}) which is related to the free evolution in the momentum representation, to obtain
\begin{eqnarray*}
\mathcal{W}_\psi( X | 0,\nu) &=& \frac{1}{|\nu|} \left|\psi\left(p, 0\right) \right|^2  \bigg|_{p=X/\nu}\,,\\[2mm]
\mathcal{W}_\psi( X | 0 ,-\nu) &=& \mathcal{W}_\psi( -X | 0 ,\nu)\,.\\
\end{eqnarray*} 
%
% FIGURE 1
\begin{figure}
\begin{center}
\includegraphics[width=0.4\linewidth]{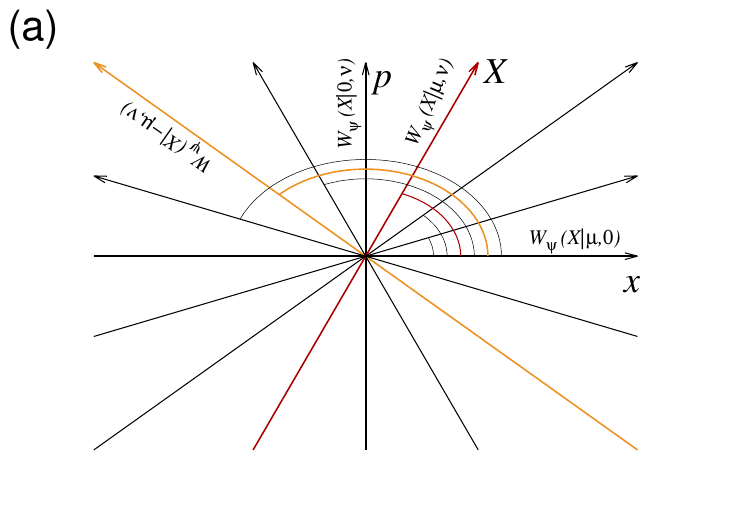}\
\includegraphics[width=0.4\linewidth]{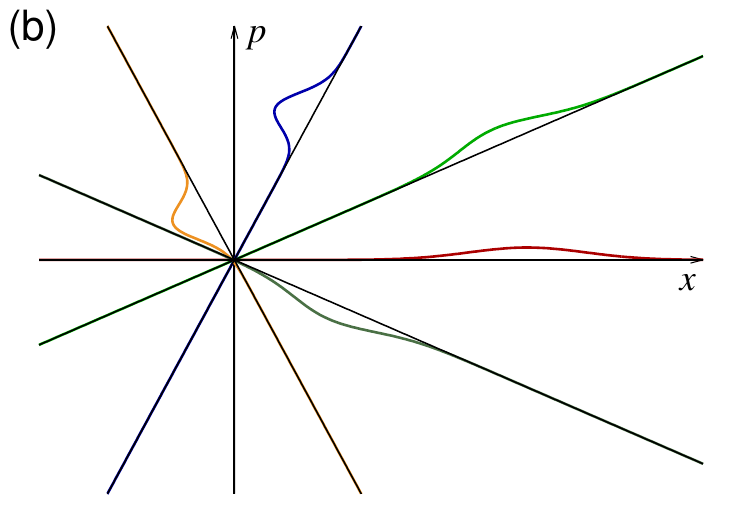}
\end{center}
\caption{(a) Shows the lines $X=\cos(\theta) x + \sin(\theta)p$ with $\theta\in[0,\pi)$ in the $xp$-plane; the arrowhead indicates the positive value $X$. (b) Sketch of what the probability density of the state would look like for each specific value of $X$.}
\label{f.map}
\end{figure}

So, the observable $X=\cos(\theta)\,x+\sin(\theta)\,p$ may be visualised as a kind of a radar representation in the  $xp$-plane (expectation value in phase space) for the tomogram of a wavepacket, where each line corresponds to coordinate $X$ for a fixed $\theta$ value; thus, as a function of the variable $\theta$, one starts at $\theta=0$ with the density probability in the position representation and moves around to the density probability in momentum representation when $\theta=\pi/2$; then in the limit $\theta\to\pi$ we have again the density probability in the position representation but with the position axis reflected. In order to maintain the function as single-valued (except at the origin) we restrict to values for $\theta\in[0,\pi)$ as shown in figure~\ref{f.map} (a). Note that this restriction does not lose information, as using the symmetry relations of the tomogram Eq.~(\ref{eq.Wpsi-full}) one may observe that for values $\theta\in[\pi,2\pi)$ the tomogram is equal to the tomogram in $\theta\in[0,\pi)$ by reflecting both, the position and momentum axes. Figure~\ref{f.map} (b) is a sketch of what the probability density of the state would look like for each specific value of $X$.

\subsection*{Time-dependent tomographic representation}

In the previous subsection we have considered the tomogram for a time-independent state $\psi(x)$, and we have shown that it may be evaluated using the free evolution operator via the identifications $\nu/\mu = \tau/m$ and $X=\mu x$, where we have used  $\tau$ to denote the free evolution-time in order to distinguished it from the natural evolution-time $t$ of the system. When we consider the time-dependent solution, for a discrete spectra of energies the expansion of the wavepacket in terms of the eigenstates of the Hamiltonian takes the form 
\begin{equation}\label{eq.Upsix}
\psi(x,t) = \sum_n c_n(t) \phi_n(x)\,; \quad c_n(t) = c_n(0) e^{-i\omega_n t}\,,
\end{equation}
and for a fixed value of time the tomogram is 
\begin{equation}\label{eq.Wpsixt}
\mathcal{W}_\psi(X,t|\mu,\nu) = \frac{1}{|\mu|} \left | \sum_n c_n(t) \varphi_n(X,\mu,\nu)\right |^2\,.
\end{equation}
We may label by $\mathcal{W}_n(X|\mu,\nu)$ the tomogram of the $n$-th eigenstate.
The particular case when $\psi(x,t)$ evolves freely yields, for $\mu,\,\nu>0$ 
\begin{equation}\label{eq.WpsixtF}
\mathcal{W}_\psi(X,t|\mu,\nu) =  \mathcal{W}_\psi\left(X\bigg|\mu,\nu+\frac{\mu}{m}t\right)\,,
\end{equation}
together with its symmetry relations~(\ref{eq.Wpsi-full}).

Equivalently, the tomogram can be written as
\begin{equation}
\mathcal{W}_\rho( X | \mu ,\nu) = \frac{1}{\vert\mu\vert} \langle X,\mu,\nu \vert \op{G}_0 \op{\rho}(t) \op{G}_0^\dagger \vert X,\mu,\nu \rangle
\end{equation}
with $\op{\rho}(t) = \op{U}(t) \op{\rho}(0) \op{U}^\dagger(t)$. The product $\op{G}_0 \op{U}(t)$ takes the role of the evolution operator for the tomogram.

For mixed states, described by
\[
\op{\rho} = \sum_a p_a \op{\rho}_a(t)
\]
the tomogram takes the form
\begin{equation}
\mathcal{W}(X;\mu,\nu) = \frac{1}{\vert\mu\vert} \sum_a p_a \langle X;\mu,\nu\vert \op{G}\op{\rho}_a(t)\op{G}^\dagger \vert X;\mu,\nu\rangle\,.
\end{equation}

\section{Examples}
\label{sec3}

\subsection{Gaussian wavepacket}

A general gaussian state centred at $x=x_0$ with standard deviation $\sigma_0$, is given by
\begin{equation}\label{eq.psiG0}
\psi(x,0;k_0) =  \left[\frac{1}{2\pi\sigma_0^2}\right]^{1/4} \exp\left[-\frac{(x-x_0)^2}{4\sigma_0^2}\right]e^{ik_0 x}\,,\quad k_0=k_0^*\,,
\end{equation}
whose probability density moves with a velocity $v_0 = \hbar k_0/m$.  The probability distribution yields 
\[\sigma_{x}^2 = \sigma_0^2\,, \quad \sigma_{p}^2 = \frac{1}{4\sigma_{x}^2}\,, \quad \sigma_{xp} = 0\,,\]
where we have defined
\[\sigma_{{\rm AB}} = \frac{1}{2}\bra \op{{\rm AB}} + \op{{\rm BA}}\ket - \bra \op{{\rm A}}\ket\bra\op{{\rm B}}\ket\,, \quad \sigma_{\rm A}^2 = \sigma_{{\rm AA}}\,,\]

Using the free particle propagator one finds the evolution of this state as 
\begin{equation}\label{eq.psiGt}
\psi(x,\tau;k_0) =  \left[\frac{1}{2\pi\sigma^2(\tau)}\right]^{1/4} \exp\left[-\frac{[x-x_0(\tau)]^2}{4\sigma^2(\tau)}\right]e^{-i[\tan^{-1}(\tau/\tau_0)+2\gamma(x,\tau)]/2}\,,
\end{equation} 
with
\begin{eqnarray*}
\tau_0 := \frac{2 m\sigma_0^2}{\hbar}\,;\qquad \sigma(\tau):=\sigma_0 \sqrt{1+\frac{\tau^2}{\tau_0^2}}\,;\qquad x_0(\tau) := x_0 + \frac{\hbar k_0}{m}\tau\,,\\
\gamma(x,\tau):= \frac{1}{4\sigma^2(\tau)}\frac{\tau_0}{\tau}\left\{[x-x_0(\tau)]^2 - \left[2x_0[x_0(\tau)-x] + x^2 - x_0^2\right]\frac{\sigma^2(\tau)}{\sigma_0^2}\right\}\,.
\end{eqnarray*}
Replacing $x\to X/\mu$, $\tau/m\to \nu/\mu$ into $\psi(x,\tau,k_0)$ provides the tomogram $\mathcal{W}_\psi( X | \mu ,\nu)$, for $\mu,\nu>0$ as
\begin{equation}\label{eq.psiGt}
W(X|\mu,\nu) =  \frac{1}{|\mu|}\left[\frac{1}{2\pi\sigma^2(\mu,\nu)}\right]^{1/2} \exp\left[-\frac{[\frac{X}{\mu}-x_0(\mu,\nu)]^2}{2\sigma^2(\mu,\nu)}\right]\,,
\end{equation}
where
\begin{eqnarray*}
\sigma(\mu,\nu)&:=&\sigma_0 \sqrt{1+\frac{\hbar^2}{4\sigma_0^4}\frac{\nu^2}{\mu^2}}\,;\qquad x_0(\mu,\nu) := x_0 + \hbar k_0\frac{\nu}{\mu}\,,
\end{eqnarray*}
By simple inspection one finds $\widetilde{\psi}(x,\tau,k_0) = \psi^*(x,\tau;-k_0)$ and we use the symmetry relations~(\ref{eq.Wpsi-full})  for real values $\mu$ and $\nu$.

%
% FIGURE 2
\begin{figure}
\begin{center}
\includegraphics[width=0.4\linewidth]{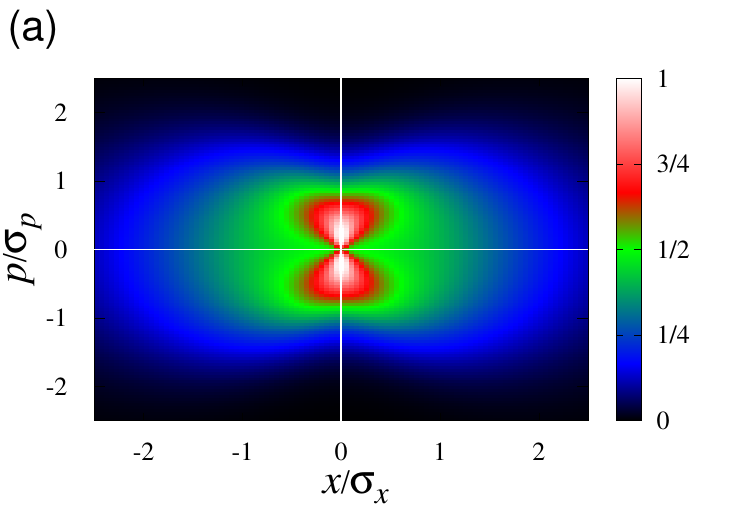}\
\includegraphics[width=0.4\linewidth]{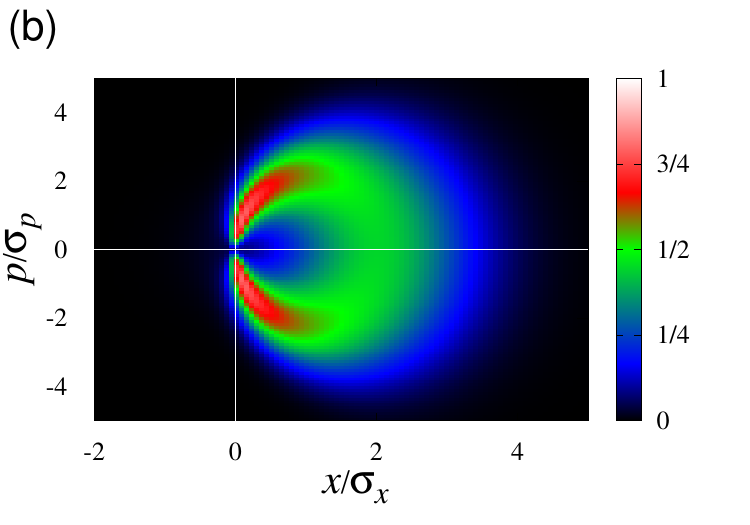} \\
\includegraphics[width=0.4\linewidth]{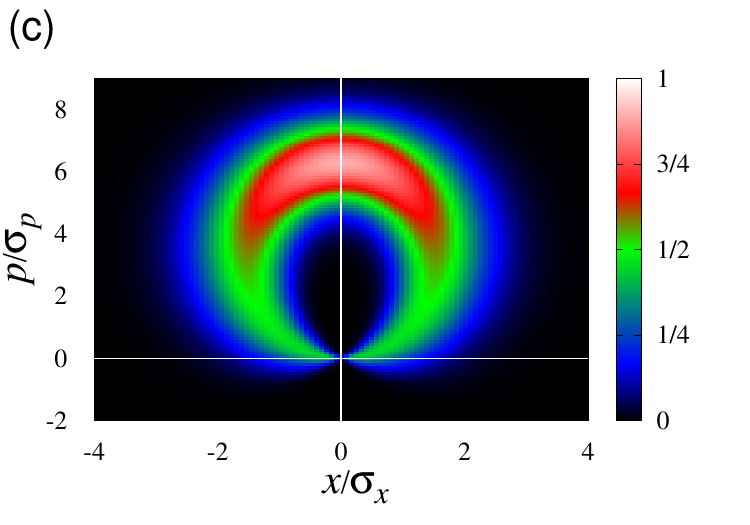}
\includegraphics[width=0.4\linewidth]{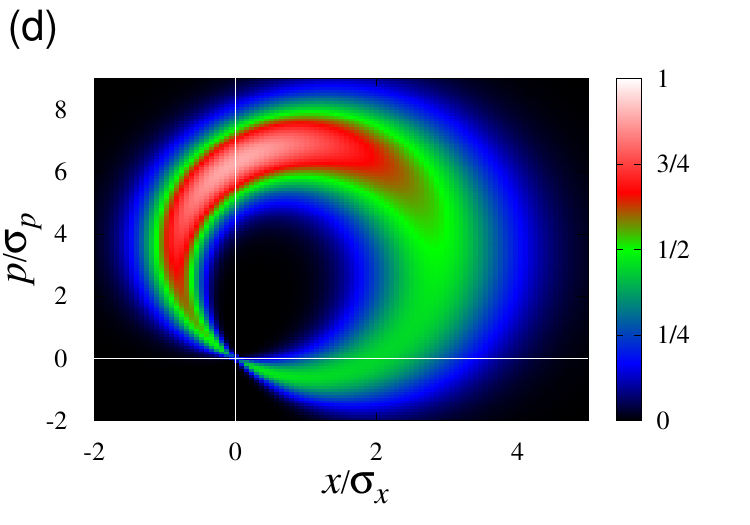}

\end{center}
\caption{Tomographic representation for a Gaussian state~(\ref{eq.psiG0}) with $\sigma_0=1$, for $\theta \in [0,\pi]$ with parameters (a) $x_0=0$, $k_0=0$;   (b) $x_0>0$, $k_0=0$;  (c) $x_0=0$, $k_0>0$ and (d) $x_0>0$, $k_0>0$; showing symmetries with respect to reflection upon the position and/or momentum axes. See text for details.}
\label{f.gaussian1}
\end{figure}

Figure~\ref{f.gaussian1} shows the tomographic representation of the gaussian wavepacket, where we have fixed $\sigma_0=1$, in the $xp$-plane, where $X=\cos(\theta)\,x+\sin(\theta)\,p$ for $\theta\in[0,\pi)$. We have taken different cases for the parameters: When centred ($x_0=0$) and static ($k_0=0$) Fig.~\ref{f.gaussian1}(a) shows a symmetric probability distribution with respect to reflection on both axes; when not centred ($x_0>0$) and static ($k_0=0$)   Fig.~\ref{f.gaussian1}(b) shows a probability distribution that is symmetric with respect to reflection on the momentum axis; when centred ($x_0=0$) but not static ($k_0>0$)  Fig.~\ref{f.gaussian1}(c) shows a probability distribution that is symmetric with respect to reflection on the position axis; finally, when not centred ($x_0>0$) and not static ($k_0>0$) Fig.~\ref{f.gaussian1}(d) shows a distribution with no symmetries. Note: for $x_0<0$ and/or $k_0<0$ the tomogram gets reflected with respect to the corresponding axis.

\subsection{Quantum shutter problem}

The quantum shutter can be modeled as a perfect absorber placed at $x=0$, which is removed at $t=0$.  A constant flux of particles comes in from the left. For $t\leq 0$ the wavefunction is described by
\begin{equation*}
\psi(x,t;k_0) = \theta(-x) e^{i (k_0 x-\omega_0 t)} \, , \quad \omega_0= \frac{\hbar k_0^2}{2m}\,,
\end{equation*}
where $\theta(x)$ is the Heaviside theta function. For $t>0$,  the wavefunction will be given in terms of the Moshinsky function $\psi(x,t;k_0) =M(x, t, k_0)$~\cite{moshinsky51,moshinsky52}. This is obtained by applying the free evolution operator to the initial state $\psi(x,0;k_0)$ as shown in~\ref{moshinsky}.

In order to distinguish the fact that the flux is incoming from the left, we will use a subscript ``L''on the function, whose explicit expression is
\begin{eqnarray}\label{eq.ML0}
M_L(x,\tau;k_0) &=& \frac{1}{2} e^{im x^2/2\hbar \tau} \omega(iy)\,, %\\[2mm] &&
\quad \textrm{with} \quad y := e^{-i\pi/4}\sqrt{\frac{m}{2\hbar \tau}} \left(x-\frac{\hbar k_0}{m}\tau\right)\,,
\end{eqnarray} 
and unnormalised initial condition
\[\Psi_L(x,0,k_0)=\lim_{\tau\to 0^+} M_L(x,\tau;k_0) = \theta(-x) e^{ik_0 x}\,.\]
Again we use $\tau$ for the free evolution as in the Green function~(\ref{eq.G0xt}), which allows us to determine the tomogram, that is, the reference frame used to determine corresponding probabilities. 
Here, $\omega(z)= e^{-z^2}{\rm erfc}\,(-iz)$ is the Faddeyeva function~\cite{abramowitzwz,faddeeva}; its relation to the Fresnel integrals is the origin of the physical phenomenon known as {\it diffraction in time} \cite{moshinsky52,brukner97,delcampo09}.

The solution above may be extended to the case when the shutter is placed at $x=x_0$ and also to the lower half of the complex $k$-plane (cf.~\ref{moshinsky}). The advantage of considering complex values is that the initial state is normalisable. Through the identification of the variables $\nu/\mu = \tau/m$ and $X/=x$ one finds the tomogram as 
\begin{equation}\label{eq.WLRsq}
\mathcal{W}_{L}(X|\mu,\nu)= \frac{1}{4|\mu|} |\omega(iy)|^2\,, \qquad y = e^{-i\pi/4}\sqrt{\frac{\mu}{2\hbar \nu}} \left(\frac{X}{\mu}-\hbar\, k\,\frac{\nu}{\mu}\right)\,,\quad {\rm Im}[k]\leq 0\,,
\end{equation}
which for real values of $k$ may be written in terms of Fresnel integrals~\cite{delcampo09}.

\begin{figure}
\begin{center}
\includegraphics[width=0.4\linewidth]{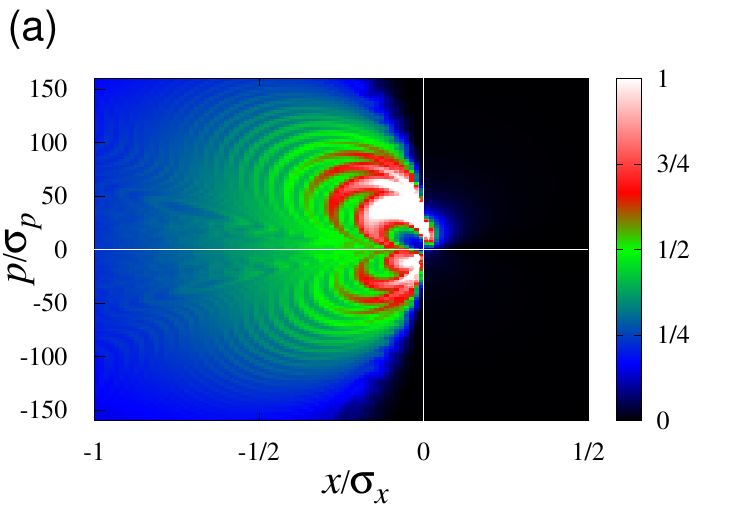}\
\includegraphics[width=0.4\linewidth]{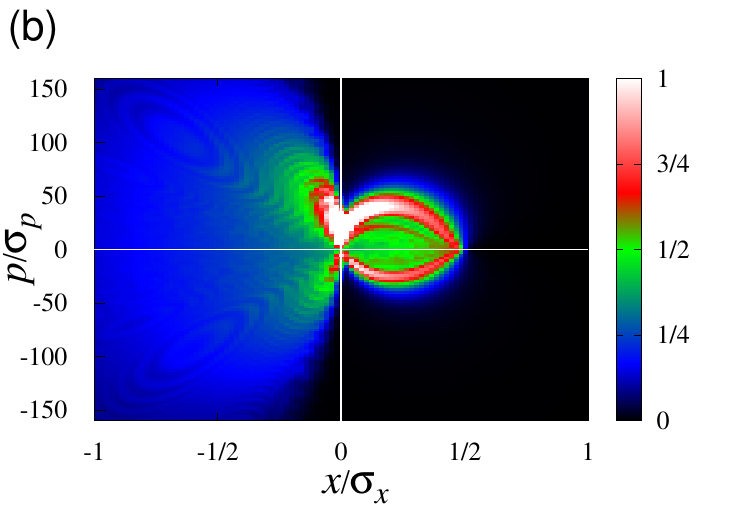}
\end{center}
\caption{Behaviour of the tomographic representation for the quantum shutter problem with incidence from the left, for complex $k=1-0.05i$, when the shutter is placed at (a) $x_0=0$ and (b) $x_0 = \sigma_x/2=5$. See text for details.}\label{f.shutter1}
\end{figure}

In order to show the tomogram of this kind of initial states for incidence from the left, we consider an unnormalised initial state of the form
\[\Psi_L(x,0;k)=\theta(x) e^{ikx}\,, \qquad k = k_r - i k_i\,,  \qquad k_i>0\,,\] 
with 
\[\sigma_{x}^2= \frac{1}{4k_i^2}\,,\qquad \sigma_{p}^2:=\frac{1}{4\sigma_{x}^2}\,, \qquad\sigma_{xp}=2k_r\left(x_0- \sigma_{x}\right)\,,\]
where $\sigma_{p}^2$ was calculated using the criterion for a Lorentzian distribution (half-width at half-maximum) for the Fourier transform of the initial state, since this distribution has no finite moments $\bra \op{p}^s\ket$ for $s=2,\,3\,,\ldots$ Notice that this state, with the convention for the value of $\sigma_p$, satisfies $2\sigma_{x}\sigma_{p} =1$ as a Gaussian distribution.

Figure~\ref{f.shutter1} shows the tomogram of this state in $xp$-space for complex $k=1-0.05i$ when the shutter is placed at $x_0=0$ [fig.~\ref{f.shutter1}(a)] and when it is placed at $x_0=\sigma_x/2$ [fig.~\ref{f.shutter1}(b)]. It is not symmetric with respect to reflection on the $p$-axis because the Fourier transform of the initial state is also not symmetric for $k_r>0$. The oscillations that the density probability suffers reflect the phenomenon named {\it diffraction in time}, found in this kind of states when they evolve.

\begin{figure}
\centering
\includegraphics[scale=0.35]{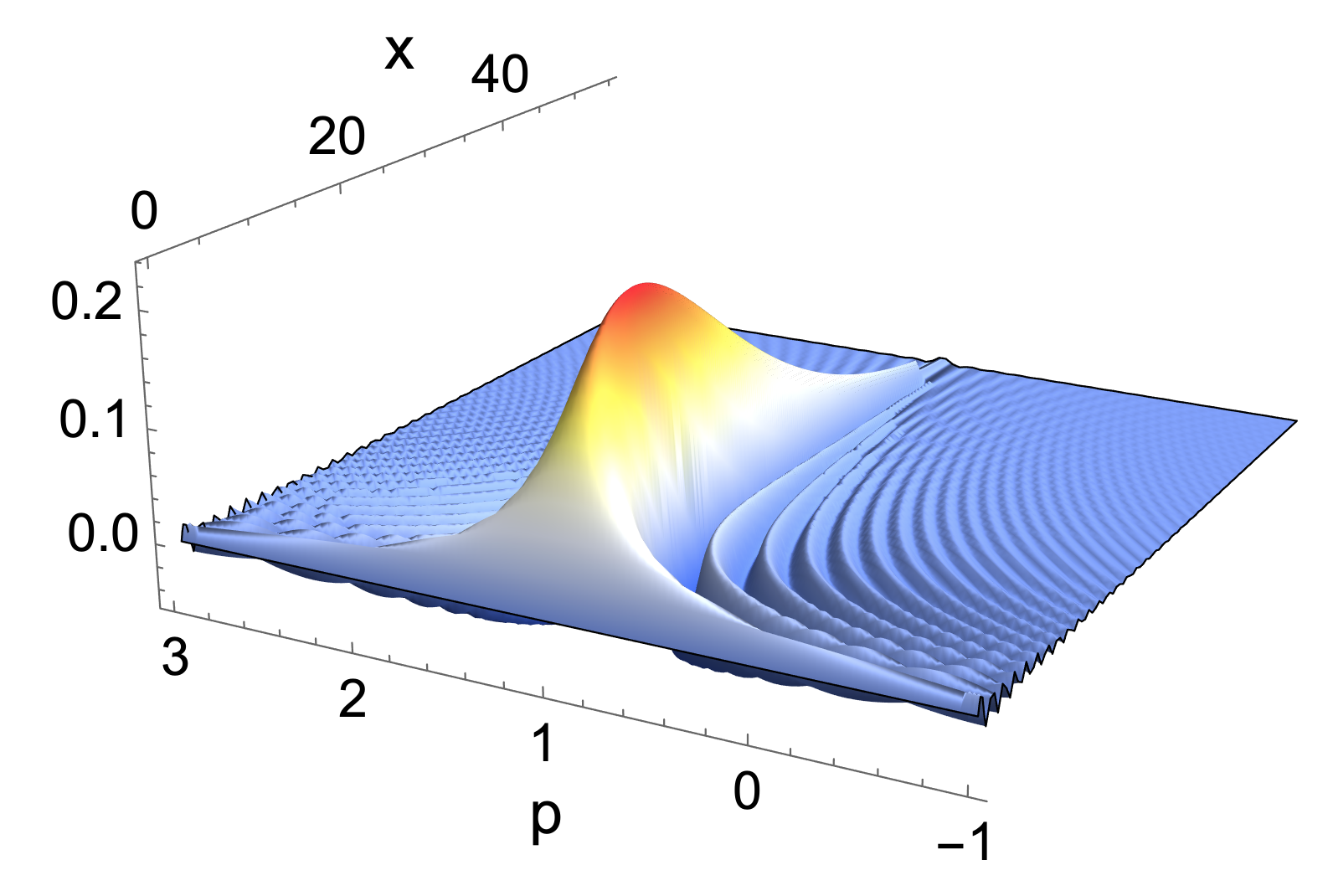}
\includegraphics[scale=0.25]{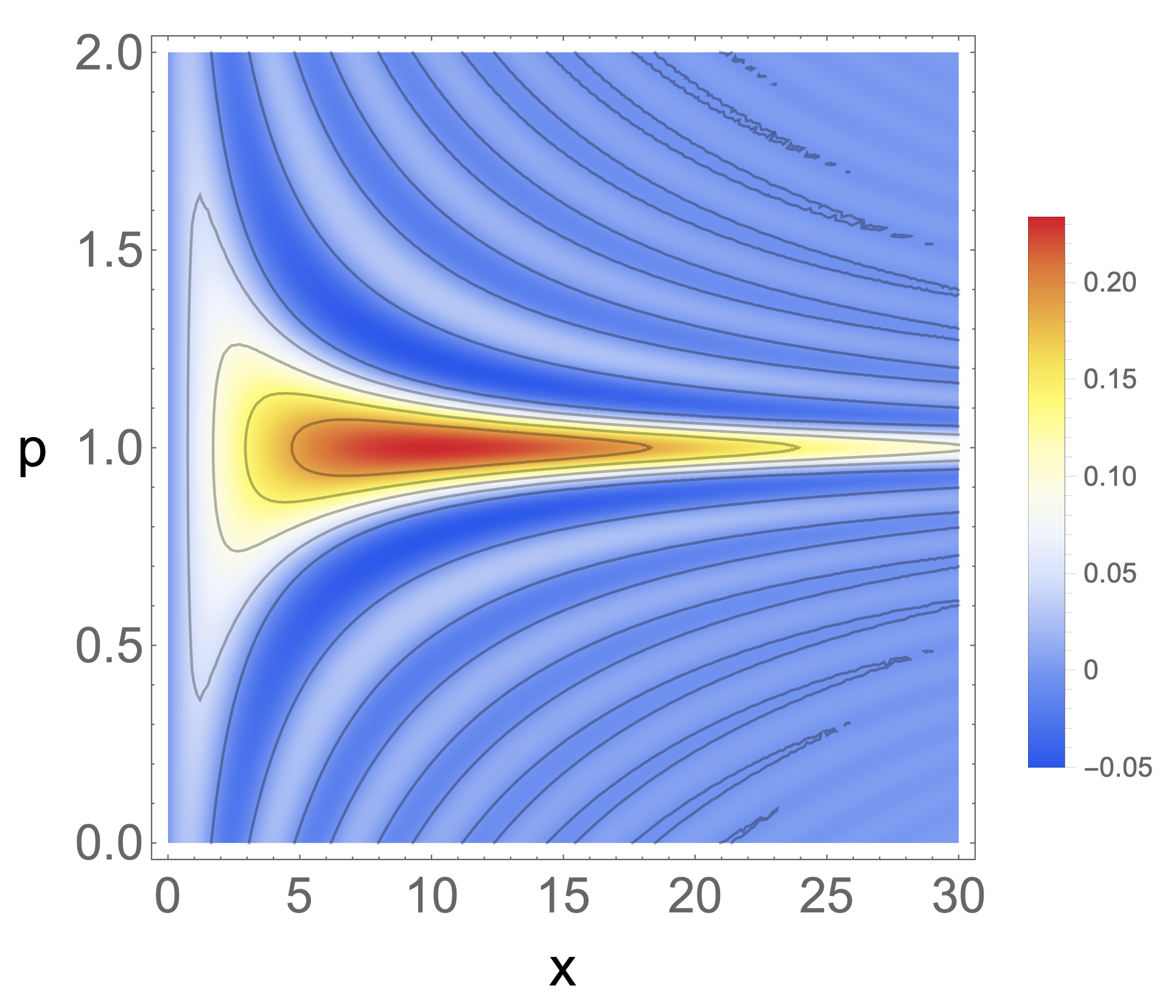}
\caption{Left: Normalized Wigner function for the state $\theta(x) e^{i k x}$ for $k=1+0.05i$.  Right: contour plot of the same, zoomed in near the peak. Negative values of the Wigner function is a signature of the quantumness of the system. \label{wignerf}}
\end{figure}

The Wigner function for the quantum shutter can also be obtained. By taking the state $\psi(x)=\theta(\pm x)e^{i k_{\pm} x}$, the normalized Wigner function of the system can be written as
\begin{equation}
W(x,p)=\pm \frac{2k_i \theta (\pm x) e^{\mp 2 k_i x} \sin (2 x (k_r-p))}{\pi  (k_r-p)}, \quad k_{\pm}=k_r \pm i k_i, \quad k_i > 0,
\end{equation}
whose plot can be see in fig.~\ref{wignerf}. In this figure, one can see the dispersion of the system as a function of $x$ and $p$. The maximum of the Wigner function is attained at  $p=k_r$ for $x\neq 0$. Ripples of much smaller amplitude are present.

This initial state, together with the solution for incidence from the right, denoted by $\Psi_R(x,\tau;k)$, permits us to calculate the tomogram via the free evolution of a set of initial states by considering the problem as the single, double or multiple shutter problem as will be shown below.

\subsection*{Double quantum shutter setup:}

For simplicity consider the unnormalised initial state
\begin{equation}\label{eq.phi0}
\phi(x,0;\kappa) = e^{i\kappa x}\left[\theta(x-a)-\theta(x-b)\right]\,,\quad a<b\,,
\end{equation}
representing a state extended along the interval $x\in[a,b]$; here $\kappa=k_r+ik_i$ is a complex number. For this state we have, with $\ell=b-a$,
\[\sigma_{x}^2=\frac{1}{4} \left[\frac{1}{k_i^2} - \ell^2{\rm csch}^2(k_i\,\ell)\right] \,,\]
which reduces to  $\sigma_{x}^2=\ell^2/12$ when  $k_i\to 0$ and assuming $2\sigma_p\sigma_x=1$, as for the Lorentzian distribution.

For the case ${\rm Im}\,(\kappa)< 0$ one rewrites $\theta(x-a)-\theta(x-b) = \theta(b-x)-\theta(a-x)$, and hence the time dependent solutions is
\[\phi_L(x,\tau;\kappa) = \Psi_L(x,\tau;b,\kappa) - \Psi_L(x,\tau;a,\kappa)\,, \quad \textrm{when}\quad {\rm Im}\,[\kappa]< 0\,;\]
and, on the other hand
\[\phi_R(x,\tau;\kappa) = \Psi_R(x,\tau;a,\kappa) - \Psi_R(x,\tau;b,\kappa)\,, \quad  \textrm{when}\quad {\rm Im}\,(\kappa)> 0\,.\]
When $\kappa=\kappa^*=k_0$ is real, one uses the symmetry relationship of the Moshinsky function given in~\ref{moshinsky} [Eq.~(\ref{eq.MLpMR})] to show that $\phi(x,\tau;k_0):=\phi_L(x,\tau;k_0) = \phi_R(x,\tau;k_0)$. A linear combination of these solutions provides in a natural way other initial states of interest; in particular the set of the Fourier basis, and hence any truncated initial state may be expanded in a Fourier series providing a tomogram in the form~(\ref{eq.Wpsixt}).

Thus, the tomogram reads
\begin{eqnarray}
&& \mathcal{W}_{L,R}(X|\mu,\nu)= \frac{1}{4|\mu|} \left|e^{ik_0b} e^{i\mu (X/\mu-b)^2/2\hbar \nu}\omega(\pm iy_b) - e^{ik_0a} e^{i\mu (X/\mu-a)^2/2\hbar \nu}\omega(\pm iy_a)\right|^2, \\[2mm]
 && y_{x_0} := e^{-i\pi/4}\sqrt{\frac{\mu}{2\hbar \nu}} \left[\frac{X}{\mu}- \left(x_0+\hbar\, k_0\,\frac{\nu}{\mu}\right)\right]\,.\nonumber
\end{eqnarray}
\begin{figure}
\begin{center}
\includegraphics[width=0.4\linewidth]{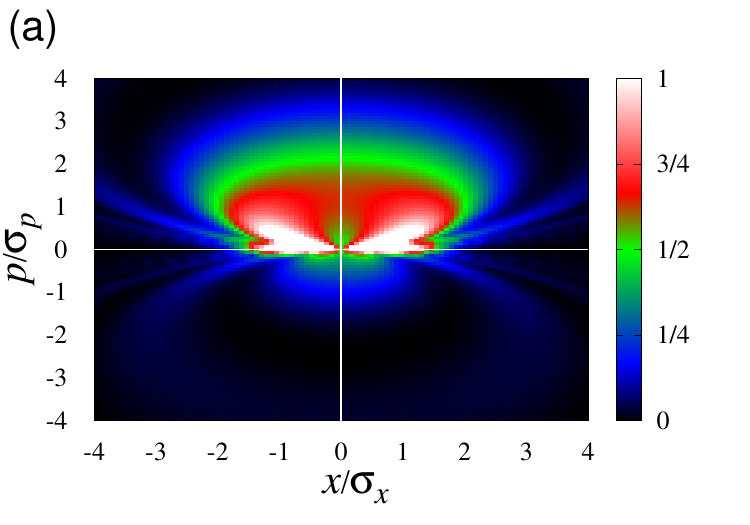}\
\includegraphics[width=0.4\linewidth]{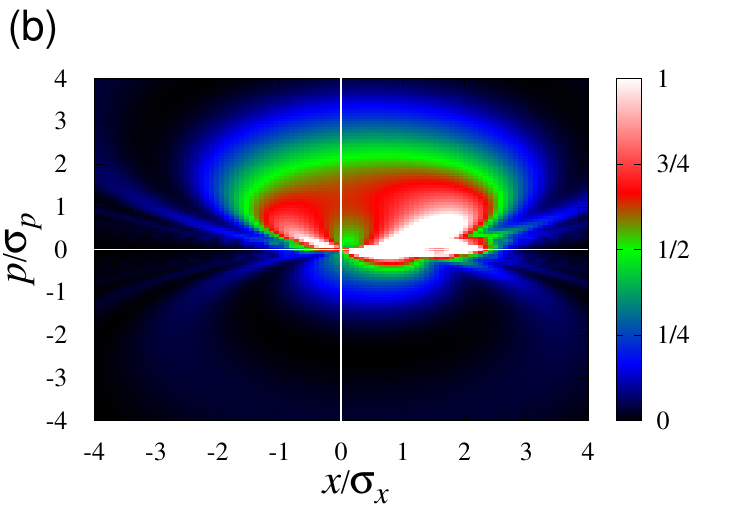}
\end{center}
\caption{Behaviour of the tomographic representation of the double quantum shutter problem  for the unnormalised initial state~(\ref{eq.phi0}) with real $\kappa=1$ and $b-a=\sqrt{12}\sigma_x=2$, for (a) the symmetric state $b=-a=1$ and (b) the non-symmetric state $b=-3a=3/2$. }\label{f.shutter2}
\end{figure}

Figure~\ref{f.shutter2} shows the behaviour of the tomogram for the initial state~(\ref{eq.phi0}), where $\kappa=1$ and $b-a=\sqrt{12}\sigma_x=2$, for both the symmetric state $b=-a=1$ [Fig.~\ref{f.shutter2}(a)] showing a symmetric probability density respect to a reflection on the $x$-axis, and the non-symmetric state $b=-3a=3/2$ [Fig.~\ref{f.shutter2}(b)].

\subsection*{States of the finite cutoff potential:}

Fixing $V(x)=0$ for $x\not\in[a,b]$ divides the space in a natural way  into three regions: (I) for values $x<a$ (or left-side region), (II) for values $x\in[a,b]$ (or internal region) and (III) for values $x>b$ (or right-side region).  The free-evolution (tomogram in our context) of an eigenstate may be seen as three shutter problems. For incidence from left to right and for incidence from right to left the solution in regions I and III when the initial state is an eigenstate of the Hamiltonian ``with complex $\kappa$ values for bound states'' are given by Moshinsky functions (left and right). The evolution in the internal region depends on the nature of the potential. However, when this is given as a multibarrier potential, each region with constant potential may be seen as a double shutter setup.  An example of this fact is to consider a simple barrier (or well) potential, i.e., $V(x) = v_0\left[ \theta(b-x)-\theta(a-x)\right]$; the unnormalised eigenfuntions of this problem are the scattering states which, for incidence from the left, take the form
\[\chi_L(x;k) = \left\{\begin{array}{l l}
e^{ik x}  +r(k) \,e^{-ikx} & x\leq  a \\[2mm]
A(k) \,e^{iq x}  +B(k) \,e^{-iqx} & a< x<b \\[2mm]
t(k)\, e^{ikx} & b \leq x
\end{array}\right.\,,\]
where $k^2 = q^2 + 2 mv_0/\hbar^2$ and $r(k)$ and $t(k)$ are the reflexion and transmission amplitudes. (A similar structure have the scattering states $\chi_R(x;k)$ for incidence from the right side.) The free evolution of this state is written as
\[\chi_L(x,\tau;k) = \chi_L^{({\rm I})}(x,\tau;k) + \chi_L^{({\rm II})}(x,\tau;k) + \chi_L^{({\rm III})}(x,\tau;k)\,,\]
with
\begin{eqnarray*}
\chi_L^{({\rm I})}(x,\tau;k) &=& \Psi_L(x,\tau;a,k) + r(k) \Psi_L(x,\tau;a,-k)\,,\\[2mm]
\chi_L^{({\rm II})}(x,\tau;k) &=& A(k) \phi_{ab}(x,\tau;q) + B(k) \phi_{ab}(x,\tau;-q)\,, \quad k^2 = q^2 + 2 mv_0/\hbar^2\\[2mm]
\chi_L^{({\rm III})}(x,\tau;k) &=& t(k) \Psi_R(x,\tau;b,k)\,,
\end{eqnarray*}
where $\phi_{ab}(x,\tau;q)$ is the appropriate free-time dependent solution of the double quantum shutter $\phi_L(x,\tau;\kappa)$ and/or $\phi_R(x,\tau;\kappa)$ depending of the value of $q$, since $k$ is real but $q$ may be real or imaginary.  So the tomogram for the scattering states from left is calculated as
\begin{equation}\label{Wscattering}
\mathcal{W}_L(X|\mu,\nu) = \frac{1}{|\mu|} \left|\chi_L\left(\frac{X}{\mu},\frac{m\nu}{\nu};k\right)\right|^2\,.
\end{equation}

The bound states, when $v_0<0$, may be obtained using the expressions for scattering states via an appropriate limit to a pole of the transmission amplitude along the imaginary axis in the lower half of the complex $k$-plane. 

One extends this kind of solutions to eigenestates of a multibarrier potential by using in each internal region with constant potential a double quantum shutter.

\section{Generalisation to $N$-particles}
\label{sec4}

Consider first the case of two modes of the harmonic oscillator without interaction. The basis of the Hilbert space is written as
\[|n_1\, n_2\ket = |n_1\ket\otimes|n_2\ket\,,\]
which are the eigenstates of the Hamiltonian $\op{H}=\op{H}_1+\op{H}_2$ with $[\op{H}_1,\op{H}_2]=0$, and an arbitrary state is given as linear combination of these states. In the position representation we denote the element of the basis as $\Xi_{n_1n_2}:=\bra x_1\, x_2|n_1\, n_2\ket = \psi_{n_1}\psi_{n_2}$, thus the tomogram of this state is~\cite{arkhipov03, chernega23} ($\hbar =1$)
\begin{equation}\label{eq.Wxmunu2}
\mathcal{W}_{n_1n_2}( X_1,X_2 | \mu_1,\mu_2 ,\nu_1,\nu_2) = \frac{1}{4 \, \pi^2 |\nu_1\nu_2|} \Bigg| \int^\infty_{-\infty}\, \Xi_{n_1n_2} \, e^{i \frac{\mu_1 x_1^2}{2  \nu_1} - i \frac{X_1 x_1}{ \nu_1}+i \frac{\mu_2 x_2^2}{2  \nu_2} - i \frac{X_2 x_2}{ \nu_2}} {\rm d} x_1{\rm d} x_2 \Bigg|^2  \, .
\end{equation}
As in the case of a single particle, one identifies for the Green function~(\ref{eq.G0xt}) the variables
\begin{equation}
\frac{\nu_s}{\mu_s} = \frac{\tau}{m_s}\,,\qquad \frac{X_s}{\mu_s}=x_s\,, \quad s=1,\,2\,,
\end{equation}
and taking account that when the state is separable the tomogram (\ref{eq.Wxmunu2}) is given by the product of the tomogram for each part, for such case one has
\begin{equation}\label{eq.Wxmunu2b}
\mathcal{W}_{n_1n_2}\left( X_1,X_2 | \mu_1,\mu_2 ,\nu_1,\nu_2\right) = \mathcal{W}_{n_1}\left( X_1 | \mu_1,\nu_1\right)\mathcal{W}_{n_2}\left(X_2 |\mu_2 ,\nu_2\right)  \, ,
\end{equation}
where
\begin{equation*}
\mathcal{W}_{n_s}(X_s|\mu_s,\nu_s) = \frac{1}{|\mu_s|}\left|\varphi_{n_s}\left(X_s|\mu_s,\nu_s\right)\right|^2\,.
\end{equation*}
Using the position representation for the Green function
\begin{equation}\label{eq.Gs}
G_s(X_s,y_s|\mu_s,\nu_s)=\sqrt{\frac{\mu_s}{2\pi i\, \nu_s}} \exp\left[i\frac{ \mu_s}{2 \nu_s} \left(\frac{X_s}{\mu_s}-y_s\right)^2\right]\,.
\end{equation}
the transformed function takes the form
\begin{eqnarray}\label{eq.varphink}
\varphi_{n_s}\left(X_s|\mu_s,\nu_s\right)&=&\int^\infty_{-\infty}\, \psi_{n_s}(y) G_s(X_s,y|\mu_s,\nu_s) {\rm d} y\nonumber \\[2mm]
&=& \frac{1}{\sqrt{2^{n_s} n_s!} \,\pi^{1/4}} \sum_{j=0}^{n_s} H_{n_s}^{(j)}(0) \widetilde{\psi}_{j}(X_s|\mu_s,\nu_s)\,,  \quad H_{n_s}^{(j)}(0) := \frac{1}{j!}\frac{{\rm d}^j}{{\rm d}y^j}H_{n_s}(y)\bigg|_{y=0}\,,
\end{eqnarray}
where $H_{n_s}^{(j)}(0)$ are the coefficients of the Hermite polynomial of order $n_{s}$, and where 
\begin{eqnarray}
\widetilde{\psi}_j(X_s|\mu_s,\nu_s) &:=& \int y^j e^{-y^2/2}G_s(X_s|\mu_s,\nu_s){\rm d} y \nonumber \\[2mm]
&=& \sqrt{\frac{2^{j+1}}{\pi}} \sqrt{\left(\frac{\nu_s}{\nu_s-i\mu_s}\right)^j\left(\frac{\mu_s}{i\nu_s}\right)}\, g_j(X_s|\mu_s,\nu_s)\,e^{iX_s^2/2\mu_s\nu_s}
\end{eqnarray}
with
\begin{eqnarray*}
g_j(X_s|\mu_s,\nu_s) &:=& \left\{ \begin{array}{l l} 
\Gamma\left(\frac{j+1}{2}\right) \sqrt{\frac{\nu_s}{2(\nu_s-i\mu_s)}}\,_1F_1\left(\frac{j+1}{2},\frac{1}{2},-\frac{i}{2}\frac{X_s^2}{\mu_s\nu_s+i\nu_s^2}\right)\,, & \textrm{even}\  j \\[4mm]
\Gamma\left(\frac{j}{2}+1\right) \frac{X_s}{\mu_s+i\nu_s}\,_1F_1\left(\frac{j}{2}+1,\frac{3}{2},-\frac{i}{2}\frac{X_s^2}{\mu_s\nu_s+i\nu_s^2}\right)\,, & \textrm{odd}\  j \end{array}\right.\,,
\end{eqnarray*}
Particular examples of these are 
\begin{eqnarray*}
\varphi_{0}\left(X_s|\mu_s,\nu_s\right) &=&  \frac{1}{\pi^{1/4}}\sqrt{\frac{\mu_s}{\mu_s+i\nu_s}} \exp\left(-\frac{1}{2}\frac{X_s^2}{\mu_s^2+i\mu_s\nu_s}\right)\,,\\[3mm]
\varphi_{1}\left(X_s|\mu_s,\nu_s\right) &=&  \frac{\sqrt{2} X_s}{\mu_s+i\nu_s}\varphi_{0}\left(X_s|\mu_s,\nu_s\right)\,,\\[2mm]
\varphi_{2}\left(X_s|\mu_s,\nu_s\right) &=&  \frac{2X_s^2-(\mu_s^2+\nu_s^2)}{\sqrt{2}(\mu_s+i\nu_s)^2}\varphi_{0}\left(X_s|\mu_s,\nu_s\right)\,,\\[2mm]
\varphi_{3}\left(X_s|\mu_s,\nu_s\right) &=&  \frac{2X_s^3-3X_s(\mu_s^2+\nu_s^2)}{\sqrt{3}(\mu_s+i\nu_s)^3}\varphi_{0}\left(X_s|\mu_s,\nu_s\right)\,,.
\end{eqnarray*}
In general one finds that
\begin{equation}
\varphi_j\left(X_s|\mu_s,\nu_s\right) =\frac{1}{\sqrt{2^jj!}} \left(\frac{\mu_s-i\nu_s}{\mu_s+i\nu_s}\right)^{j/2} H_j\left(\frac{X_s}{\sqrt{\mu_s^2+\nu_s^2}}\right)\varphi_0\left(X_s|\mu_s,\nu_s\right) \,,
\end{equation}
exhibits the solution as a polynomial of order $n_s$ plus the transformed wave function of the vacuum state.
For a general state which is given as a linear combination of the basis states the tomogram takes the form
\begin{equation*}
\mathcal{W}_\psi(X_1,X_2|\mu_1,\mu_2,\nu_1,\nu_2) = \frac{1}{|\mu_1\mu_2|}\left|\sum_{n_1,n_2} a_{n_1n_2}\varphi_{n_1}\left(X_1|\mu_1,\nu_1\right)\varphi_{n_2}\left(X_2|\mu_2,\nu_2\right)\right|^2\,,
\end{equation*}
and since the particles are not interacting their evolution is simple and the time-dependent tomogram reads as
\begin{equation*}
\mathcal{W}_\psi(X_1,X_2,t|\mu_1,\mu_2,\nu_1,\nu_2) = \frac{1}{|\mu_1\mu_2|}\left|\sum_{n_1,n_2} a_{n_1n_2}e^{-i \varepsilon_{n_1n_2}t}\varphi_{n_1}\left(X_1|\mu_1,\nu_1\right)\varphi_{n_2}\left(X_2|\mu_2,\nu_2\right)\right|^2\,,
\end{equation*}
where $\varepsilon_{n_1n_2} = \varepsilon_{n_1}+\varepsilon_{n_2}$ are the eigenvalues of the Hamiltonian. As pointed out above, when the state is separable the tomogram is the product of two tomograms, while for a non-separable state the tomogram in general has the contribution of the interference between the transformed wave functions. So, for the case of a simple symmetric and antisymmetric initial state,
\[
|jk\ket_\pm = \frac{1}{\sqrt{2}}\left[|j\ket\otimes|k\ket \pm |k\ket\otimes|j\ket\right]\,, \quad j\neq k\,,
\]
the time-dependent tomogram is
\begin{eqnarray}\label{eq.Wpm}
\mathcal{W}_\pm(X_1,X_2,t|\mu_1,\mu_2,\nu_1,\nu_2) &=& \frac{1}{2} \left[\mathcal{W}_{j}^{(1)}\mathcal{W}_{k}^{(2)} + \mathcal{W}_{k}^{(1)}\mathcal{W}_{j}^{(2)} \right] \nonumber \\[3mm]
&& \pm \frac{1}{2|\mu_1\mu_2|}\left[\varphi_{j}^{(1)}(t)\varphi_{k}^{(2)}(t) \left(\varphi_{k}^{(1)}(t)\varphi_{j}^{(2)}(t)\right)^*+\textrm{c.c.}\right]\,,
\end{eqnarray}
where we have simplified the notation by setting $\mathcal{W}_j^{(s)} = \mathcal{W}_j(X_s|\mu_s,\nu_s)$ and $\varphi_{j}^{(s)}(t) = e^{-i\varepsilon_s t}\varphi_{j}(X_s|\mu_s,\nu_s)$. One observes that, if $\varepsilon_{jk} = \varepsilon_{kj}$ then the tomogram of the initial state $|jk\ket_\pm$ remains constant under time evolution. 

Integration with respect to one variable $X_{s'}$ yields the tomogram of the reduced density matrix of the $s$-particle, and using the fact that the states are ortogonal one finds
\begin{equation*}
\mathcal{W}_\pm^{(s)}(X_s|\mu_s,\nu_s) = \frac{1}{2} \left[\mathcal{W}_{j}^{(s)} + \mathcal{W}_{k}^{(s)} \right] \,,
\end{equation*}
which is independent of time and is the same for the two states.

The expressions above may be generalised to an initial state of the form
\begin{eqnarray*}
|\Psi_{jk}\ket_\pm = a |\alpha_j\ket\otimes|\alpha_k\ket \pm b|\alpha_k\ket\otimes|\alpha_j\ket\,,
\end{eqnarray*}
where $|\alpha_j\ket\neq |\alpha_k\ket$ are arbitrary states; this provides the tomogram in the form~(\ref{eq.Wpm}) where now $\varphi_{j}^{(s)}(t) = \bra X_s;\mu_s,\nu_s|\op{G}_s\op{U}_s|\alpha_j\ket$ with $\op{U}_s=e^{-i\op{H}_st}$, and that corresponding to the reduced system reads
\begin{eqnarray}\label{eq.Wreduced}
\mathcal{W}^{(s)}_{\pm}(X_s,t|\mu_s,\nu_s) &=& \left| a\right|^2 \mathcal{W}_j^{(s)}(t) + \left| b\right|^2 \mathcal{W}_k^{(s)}(t) \nonumber \\[2mm]
&&  \pm \frac{1}{|\mu_s|} \left[a b^*  \varphi_j^{(s)}(t) \left(\varphi_k^{(s)}(t)\right)^* \bra \alpha_{j}^{(s')}|\alpha_{k}^{(s')}\ket  +   \textrm{c.c.}\right] \,,
\end{eqnarray}
which reduces to the previous case when $|a|^2=|b|^2$ and $\bra \alpha_{j}^{(s')}|\alpha_{k}^{(s')}\ket=\delta_{jk}$.
  
The expression of the tomogram for two particles may be generalised to $N$ particles, by taking the Green function of a free particle as an operator $\op{G}$, which is is different from the evolution operator of the system $\op{U}(t) = e^{-i\op{H}t}$; in terms of $\op{G}$ the transformed wave function reads 
\begin{equation}\label{eq.varphiG}
\varphi\left(X|\mu, \nu\right)=\bra X;\mu,\nu|\op{G}|\psi\ket \,,
\end{equation}
where $|\psi\ket$ is the state of the free particle system. For a system of $N$ particles the general state of the system $|\vec{\Psi}\ket$ is written as a linear combination of the product of desirable states
\[|\vec{\Psi}\ket = \sum_\ell a_\ell |\vec{\psi}_\ell\ket\,; \qquad |\vec{\psi}_\ell\ket = |\psi^{(1)}_\ell\ket\otimes|\psi^{(2)}_\ell\ket\otimes\cdots|\psi^{(N)}_\ell\ket\,, \quad \bra \vec{\Psi}|\vec{\Psi}\ket = 1\,,\]
(here $|\psi_\ell^{(k)}\ket$ denotes the state of the $k$-th particle). The tomogram operator is then given by the product $\op{G} = \op{G}_1\otimes\op{G}_2\otimes\cdots \otimes\op{G}_N$ and hence one finds the transformed wave function as
\begin{equation*}
\varphi\left(\vec{X}|\vec{\mu}, \vec{\nu}\right)=\bra \vec{X};\vec{\mu},\vec{\nu}|\op{G}|\vec{\Psi}\ket = \sum_{\ell} a_\ell \prod_{s=1}^N\varphi_{\ell}^{(s)}\left(X_s|\mu_s,\nu_s\right) \,,
\end{equation*}
where $\varphi_\ell^{(s)}\left(X_s|\mu_s,\nu_s\right)$ stands for the transformed wave function $\psi^{(s)}_\ell(x)$ of the $s$-th particle, as in previous examples, and where $\vec{X} = (X_1,X_2,\dots,X_N),\ \vec{\mu} = (\mu_1,\mu_2,\dots,\mu_N)$ and $\vec{\nu} = (\nu_1,\nu_2,\dots,\nu_N)$ are used to simplify the notation. We note that, in general, $[\op{G},\op{U}(t)]\neq 0$, except when the system evolves freely, thus the time-dependent tomogram should be calculated for a fixed value of $t$, i.e., 
\begin{equation*}
\varphi\left(\vec{X},t|\vec{\mu}, \vec{\nu}\right)=\bra \vec{X};\vec{\mu},\vec{\nu}|\op{G}|\vec{\Psi}(t)\ket\,.
\end{equation*}
The form is the same that in the stationary case~(\ref{eq.varphiG}) and hence  to find the time-dependent tomogram from the expressions of the stationary one is straightforward.

Since the scale factor that appears in the tomogram depends of the representation used in the transformation of the wave function, Eqs.~(\ref{eq.Wxmunu}) and (\ref{eq.Wpnumu}), one may write the tomogam, without loss of generality, as
\begin{equation}\label{eq.WpsimunuNp}
\mathcal{W}_{\vec{\Psi}}\left( \vec{X} | \vec{\mu} ,\vec{\nu}\right) = \frac{1}{|\gamma|}\Bigg|\sum_\ell   a_\ell \varphi(\vec{X}|\vec{\mu}, \vec{\nu}) \Bigg|^2 \, , \quad \gamma = \prod_{s=1}^N \mu_s\neq 0\,,
\end{equation}
where the position representation is used for all states. The cases when $\gamma=0$ may be evaluated with an appropriate limit as mentioned in Sec.~\ref{s.freeU}.

\section{Accesible information through the tomographic picture}
\label{sec5}

For a bipartite pure system a measurement of the entanglement is the Von Neumann entropy 
$S~=~-{\rm Tr}[\op{\rho}\ln\op{\rho}]$, where $\op{\rho}$ stands for a reduced density matrix of the full system. Expanding around the identity one has 
\begin{equation}
S = S_L + \sum_{j=2}^\infty \frac{1}{j} {\rm Tr}\left[\op{\rho}\left(\op{1}-\op{\rho}\right)^j\right]\,,\qquad S_L = 1 - {\rm Tr}[\op{\rho}^2]\,,
\end{equation}
where $S_L$ is the linear entropy, so the Von Neumann entropy is equal to the linear entropy plus a residual contribution. Since $\op{1}-\op{\rho}$ is a positive matrix each term of the residual contribution takes the values zero (when $S_L=0$) or positive (when $S_L>0$) and hence $S\geq S_L$ indicating that both are a measurement of entanglement, with equality valid only for pure states $S=S_L=0$ or, in this context, for separable states.

As an example consider the general normalised, but not necessarily orthogonal, state of two particles of the form 
\begin{equation}\label{eq.psijklm}
|\Psi_{jk\ell m}\ket = a |\alpha_j\, \alpha_k\ket + b |\alpha_\ell\, \alpha_m\ket\,, \quad |\alpha_s\, \alpha_r\ket:=|\alpha_s\ket\otimes|\alpha_r\ket\,, \quad \bra \alpha_s|\alpha_s\ket =1\,;
\end{equation}
the full density matrix is then 
\begin{equation*}
\op{\varrho}_{jk\ell m} =  |a|^2\, \op{\rho}_{jkjk} + |b|^2\, \op{\rho}_{\ell m\ell m} + ab^* \,\op{\rho}_{jk\ell m} + a^*b\, \op{\rho}_{\ell mjk}\,, \quad {\rm Tr}[\op{\varrho}_{jk\ell m}]=1\,,
\end{equation*}
where
\[\op{\rho}_{abcd}:= |\alpha_a\,\alpha_b\ket\bra \alpha_c\,\alpha_d|\,, \]
which stands for a {\it quasi-projector} since it satisfies $\op{\rho}_{abcd}^2 = \bra \alpha_c\, \alpha_d|\alpha_a\,\alpha_b\ket \op{\rho}_{abcd}$ and when $c=a$ and $b=d$ one has $\op{\rho}_{abab}^2= \op{\rho}_{abab}$. Notice that $\op{\rho}_{abcd}$ is a traceless operator when the states are ortogonal $\bra \alpha_c\, \alpha_d|\alpha_a\,\alpha_b\ket=0$.  The reduced density matrix is obtained by performing a partial trace over one of the particles; tracing over the second particle this takes the form
\begin{equation*}
\op{\rho} = {\rm Tr}_{k m}\left[ \op{\varrho}_{jk\ell m}\right] = 
|a|^2\, \op{\rho}_{jj} + |b|^2\, \op{\rho}_{\ell\ell} + ab^*c_{km}^* \,\op{\rho}_{j\ell} + a^*b c_{km}\, \op{\rho}_{\ell j}\,, \qquad c_{rs}:=\bra \alpha_r|\alpha_s\ket\,,
\end{equation*} 
with $\op{\rho}_{rs} = |\alpha_r\ket\bra\alpha_s|$. A similar expression is obtained when we trace over the first particle. 

To calculate $S_L$ we require $\op{\rho}^2$ which reads as
\begin{eqnarray*}
\op{\rho}^2 &=& 
         \left[ |a|^4 + |a|^2|b|^2|c_{km}|^2 +  |a|^2 a b^*c_{\ell j}c_{km}^* + |a|^2 a^*bc_{\ell j}^*c_{km} \right] \op{\rho}_{jj}\\[1mm] 
&& + \left[ |b|^4 + |b|^2|a|^2|c_{km}|^2 +  |b|^2 a b^* c_{\ell j}c_{km}^* + |b|^2 a^*bc_{\ell j}^*c_{km}\right] \op{\rho}_{\ell\ell}\\[1mm]
&& + \left[ |a|^2|b|^2c_{\ell j} +|a|^2 a^* b c_{km} + |b|^2a^*b c_{km} + (a^*bc_{km})^2c_{\ell j}^* \right] \op{\rho}_{\ell j}\\[1mm]
&& + \left[ |a|^2|b|^2c_{\ell j}^* +|a|^2 a b^* c_{km}^* + |b|^2ab^* c_{km}^* + (ab^*c_{km}^*)^2c_{\ell j} \right] \op{\rho}_{j\ell}\,,
\end{eqnarray*}
and then the linear entropy takes the value
\begin{eqnarray}\label{eq.SLjl}
S_L %&=& 1 - \left[|a|^4 + |b|^4 + 2|a|^2|b|^2 \left(|c_{km}|^2+|c_{lj}|^2\right) + 2{\rm Re}\left[(ab^*c_{\ell j}c_{km}^*)^2\right] \right]\nonumber \\[2mm]
%&&  - 4\left(|a|^2+|b|^2\right){\rm Re}\left[ a b^* c_{\ell j} c_{km}^* \right]\nonumber \\[2mm]
&=& 2 |a|^2|b|^2(1-|c_{km}|^2)(1-|c_{\ell j}|^2)\,.
\end{eqnarray}
The expression above is the general expression for the entanglement of particles in the state~(\ref{eq.psijklm}).  Clearly, separable states with $S_L=0$ are obtained by choosing the parameters $(a,b)=(1,0)$ or $(a,b)=(0,1)$, or when the states satisfy $|\alpha_k\ket=|\alpha_m\ket$ or $|\alpha_\ell\ket=|\alpha_j\ket$ implying $|c_{km}|^2=1$ or $|c_{\ell j}|^2=1$ respectively. Also, the maximum entangled state with  $S_L=1/2$ is obtained when choosing ortogonal states $c_{km}=c_{\ell j}=0$ and equal probabilities $|a|^2=|b|^2=1/2$.

The tomographic picture allows us to calculate the available information of entangled subsystems: from the reduced density matrix, and using the free propagator, one may calculate the reduced tomogram
\begin{eqnarray}\label{eq.opW}
\mathcal{W}_1\left(X_1;\mu_1,\nu_1\right) &=& \frac{1}{|\mu_1|}\bra X_1;\mu_1,\nu_1|\op{G}_1\op{\rho}\,\op{G}_1^\dag|X_1;\mu_1,\nu_1\ket \nonumber \\[2mm]
&=& \frac{1}{|\mu_1|} \left[|a|^2\bra X_1;\mu_1,\nu_1|\op{G}_1\op{\rho}_{jj}\op{G}_1^\dag|X_1;\mu_1,\nu_1\ket\right.\nonumber \\[2mm]
&& + |b|^2\bra X_1;\mu_1,\nu_1|\op{G}_1\op{\rho}_{\ell\ell}\op{G}_1^\dag|X_1;\mu_1,\nu_1\ket\nonumber \\[2mm]
&& + ab^*c_{km}^*  \bra X_1;\mu_1,\nu_1|\op{G}_1\op{\rho}_{j\ell}\op{G}_1^\dag|X_1;\mu_1,\nu_1\ket\nonumber \\[2mm]
&& \left.+ a^*bc_{km} \bra X_1;\mu_1,\nu_1|\op{G}_1\op{\rho}_{\ell j}\op{G}_1^\dag|X_1;\mu_1,\nu_1\ket\right] \,,
\end{eqnarray}
in accordance with~(\ref{eq.Wpm}) for the particular case $\ell=k$ and $m=j$, which for uncoupled systems generalises to the time-dependent tomogram~(\ref{eq.Wreduced}) by replacing $\bra X_1;\mu_1,\nu_1|\op{G}_1\op{\rho}_{\ell j}\op{G}_1^\dag|X_1;\mu_1,\nu_1\ket\to  \bra X_1;\mu_1,\nu_1|\op{G}_1\op{U}_1(t)\op{\rho}_{\ell j}\op{U}_1^\dag(t)\op{G}_1^\dag|X_1;\mu_1,\nu_1\ket$. One could then associate the Shannon entropy of information theory with the one-position state using the standard relation known in classical probability theory
\begin{equation}
S(\mu_1,\,\nu_1) = -\int \mathcal{W}_1\left(X_1;\mu_1,\nu_1\right)\, \ln[\mathcal{W}_1\left(X_1;\mu_1,\nu_1\right)]\, dX_1\,.
\end{equation}
This classical entropy is introduced in a less critical manner, and when the distributions are highly concentrated in a region of $X$-space, it may take negative values~\cite{wehrl78}.

\section{Summary and concluding remarks.}
\label{sum}

We have shown that the symplectic tomogram associated to a quantum system may be obtained via the free evolution operator. The procedure is the same for the time-dependent tomogram, by distinguishing the natural system time $t$ from the free evolution-time $\tau$. This equivalence is made by comparing the kernel for the symplectic tomogram integral transform with the Green function of the free particle through the identification of the symplectic parameters $\mu$, $\nu$ with the free evolution-time and mass $m$ of the free particle, i.e., $\frac{\nu}{\mu}=\frac{\tau}{m}$. This may be extended to an $N$-particle system by considering the Green function for each constituent and making the specific identification of the reference frame parameters with those that label the particles. 

Various symplectic tomograms are obtained for different quantum systems, in order to exemplify the method. These include the Gaussian wave packet, the quantum shutter related with the phenomenon of diffraction in time, the double quantum shutter, and a finite potential. The generalisation to $N-$particles was used to illustrate the entanglement between two modes in a general, not necessarily orthogonal, system. The connection of the Shannon entropy with the information obtained through the reduced density matrix and free evolution operator was established. This allows us to study the tomograms for mixed states, as is possible through linear entropy. The same may be said for other quantities such as the  relative entropy and the mutual information, used to detect both quantum entanglement and superposition of states.

The tomographic representation of states allows us to consider a quantum system using only probabilities, as in  classical statistical mechanics systems. Since the path integral is related to the Feynman propagator of a system, one would expect a relation between the tomogram approach and  the path integral formalism to exist.

Only symplectic linear transformations of the position and momentum operators have been considered to define our tomograms. One may extend this to non-linear symplectic maps, which have been
developed for non-linear optics and to track particles through beam transport systems.

\appendix
\section{The Moshinsky Function}
\label{moshinsky}

The Moshinsky function $M(x,t;k_0)$ is the free solution of the time-dependent Schr\"odinger equation  for the quantum shutter setup with incident flux from the left. The initial condition is
\begin{equation*}
\psi_0(x,0) = e^{ik_0x}\theta(-x)\,,
\end{equation*} 
with $\theta(x)$ the step Heaviside function. To find $\psi_0(x,t)$ one uses a standard procedure considering the initial state as the limit of the function
\begin{equation*}
\psi_\epsilon(x,0) = e^{ik_\epsilon x}\theta(-x)\,, \quad \psi_0(x,0)=\lim_{\epsilon\to0^+}\psi_\epsilon(x,0)\,,
\end{equation*}
where $k_\epsilon:=k_0-i\epsilon$ with $\epsilon>0$. Thus, the free time-dependent solution $\psi_\epsilon(x,t)$ is
\begin{equation}\label{eq.psixte}
\psi_\epsilon(x,t) =  \frac{1}{\sqrt{2\pi}}\int_{-\infty}^{\infty} \widetilde{\psi}_\epsilon(k) e^{ikx -i\hbar k^2 t/2m }{\rm d}k\,, \quad \widetilde{\psi}_\epsilon(k) = \frac{i}{\sqrt{2\pi}} \frac{1}{k-k_\epsilon}\,,
\end{equation}
with $\widetilde{\psi}_\epsilon(k)$ the Fourier transform of the initial state. To solve the integral in (\ref{eq.psixte}) one manipulates the term in the exponent to write
\begin{equation*}
i\left(kx -\frac{\hbar}{2m} k^2 t\right) = -i\left(k-\frac{m x}{\hbar t}\right)^2 \frac{\hbar t}{2m} +i \frac{m x^2}{2\hbar t} =: - u^2 + i \frac{m x^2}{2\hbar t}\,,
\end{equation*}
which defines in a natural way the complex variable 
\begin{equation*}
u = e^{i\pi/4} \sqrt{\frac{\hbar t}{2m}}\left(k-\frac{m x}{\hbar t}\right)\,,
\end{equation*}
and in terms of this the integral (\ref{eq.psixte}) reads as
\begin{equation}\label{eq.psixteu}
\psi_\epsilon(x,t) = \frac{1}{2} e^{im x^2/2\hbar t} \left[\frac{i}{\pi} \int_{-\infty \sqrt{i}}^{\infty \sqrt{i}} \frac{e^{-u^2}}{iy_\epsilon+u}{\rm d}u\right]\,,\quad y_\epsilon := e^{-i\pi/4}\sqrt{\frac{m}{2\hbar t}} \left(x-\frac{\hbar k_\epsilon}{m}t\right)\,.
\end{equation}
The term inside brackets is related to the integral form of the Faddeyeva function $\omega(z)$~\cite{abramowitzwz,faddeeva} for complex argument $z$, which is defined as
\begin{equation*}
\omega(z) := e^{-z^2} \left(1+\frac{2i}{\sqrt{\pi}} \int_{0}^{z} e^{u^2}{\rm d}u\right)\,\,
\end{equation*}
with symmetry relationships 
\begin{equation*}
\omega(z)+\omega(-z) = 2 e^{-z^2}\,,\qquad \omega(z^*) = \omega^*(-z)\,,
\end{equation*}
and integral representation in the upper half complex plane 
\begin{equation*}
\omega(z) =\frac{i}{\pi}\int_{-\infty}^{\infty} \frac{e^{-v^2}}{z-v}{\rm d}v\,, \qquad {\rm Im}\,[z]> 0\,.
\end{equation*}
By changing the variable $u\to -v$ in~(\ref{eq.psixteu}) and analysing the position of the complex pole $iy_\epsilon$ with values $m>0,\, t>0$ and $\epsilon>0$, the Cauchy theorem yields
\begin{equation}\label{eq.psixteu2}
\psi_\epsilon(x,t) = \frac{1}{2} e^{im x^2/2\hbar t}\omega(iy_\epsilon)=:M(x,t;k_\epsilon)\,,\quad y_\epsilon := e^{-i\pi/4}\sqrt{\frac{m}{2\hbar t}} \left(x-\frac{\hbar k_\epsilon}{m}t\right)\,,
\end{equation}
which is well-defined in the limit $\epsilon\to0^+$ for real values of $k_0$.

As was shown, the $M(x,t;\kappa)$-function extends in a natural way to the lower half complex $\kappa$-plane with initial condition
\begin{equation*}
\lim_{t\to0^+}M(x,t;\kappa) = e^{i\kappa x}\theta(-x)\,,
\end{equation*}
since the solution represents a state with initial flux from the left. 

Similar considerations are made when the flux arrives from the right, with $\kappa$ in the upper half complex plane. In order to distinguish the solutions we denote these as $M_L(x,t;\kappa)$ and $M_R(x,t;\kappa)$ respectively. Following a similar procedure one finds 
\begin{equation*}
M_R(x,t;\kappa) =  \frac{1}{2} e^{im x^2/2\hbar \tau} \omega(-iy)\,, \quad y := e^{-i\pi/4}\sqrt{\frac{m}{2\hbar t}} \left(x-\frac{\hbar \kappa}{m}t\right)\,, \quad {\rm Im}[\kappa]\geq 0\,,
\end{equation*}
which satisfies the initial condition
\begin{equation*}
\lim_{t\to0^+}M_R(x,t;\kappa) = e^{i\kappa x}\theta(x)\,,
\end{equation*}
The above solutions are in accordance with the evolution of an eigenstate of the free particle. For real values $\kappa=k_0$ and using the symmetry relationship of the Faddeyeva function one has
\begin{equation}\label{eq.MLpMR}
M_L(x,t;k_0) + M_R(x,t;k_0) = e^{ik_0 x - i\hbar k_0^2 t/2m}\,.
\end{equation}
Noticing that the right hand side of equation~(\ref{eq.MLpMR}) does not depend on the position of the shutter, when this is placed at $x=x_0$ from~(\ref{eq.MLpMR}) one finds the solutions 
\begin{eqnarray*}
\Psi_L(x,t;x_0,k_0) &=& e^{ik_0x_0}M_L(x-x_0,t;k_0) \,,\quad \Psi_L(x,0;x_0,k_0) = \theta(x_0-x)e^{ik_0x}\,,\\[2mm]
\Psi_R(x,t;x_0,k_0) &=& e^{ik_0x_0}M_R(x-x_0,t;k_0)\,,\quad \Psi_R(x,0;x_0,k_0) = \theta(x-x_0)e^{ik_0x}\,.
\end{eqnarray*}
The  left and right solutions may be extended, respectively, to the lower half and upper half complex $\kappa$-plane, and these extensions are useful in order to find the free propagation of wave packets when these may be divided as a problem of multiple quantum shutters, such as the states of the infinite potential well.

\section*{References}

%\bibliographystyle{iopart-num}

%\bibliography{referencias}

\providecommand{\newblock}{}

\end{document}